\documentclass[twocolumn,showpacs,superscriptaddress,preprintnumbers,amsmath,amssymb,prb]{revtex4}

\usepackage{graphicx}
\usepackage{dcolumn}
\usepackage{bm}
\usepackage{ulem}
\usepackage{color}

\usepackage{epstopdf}

\begin{document}

\title{Robust wedge demonstration to optical negative index metamaterials}

\author{Nian-Hai Shen}
\email[]{nhshen@ameslab.gov} \affiliation{Ames Laboratory and Department of Physics and Astronomy, Iowa State University, Ames, Iowa 50011, U.S.A.}

\author{Thomas Koschny}
\affiliation{Ames Laboratory and Department of Physics and Astronomy, Iowa State University, Ames, Iowa 50011, U.S.A.}

\author{Maria Kafesaki}
\affiliation{Institute of Electronic Structure and Laser, FORTH, 71110 Heraklion, Crete, Greece} \affiliation{Department of Materials Science and Technology, University of Crete, 71003 Heraklion, Crete, Greece}

\author{Costas M. Soukoulis}
\affiliation{Ames Laboratory and Department of Physics and Astronomy, Iowa State University, Ames, Iowa 50011, U.S.A.}
\affiliation{Institute of Electronic Structure and Laser, FORTH, 71110 Heraklion, Crete, Greece}

\begin{abstract}

A robust wedge setup is proposed to unambiguously demonstrate negative refraction for negative index metamaterials. We applied our setup to several optical metamaterials from the literature and distinctly observed the phenomena of negative refraction. This further consolidates the reported negative-index property. It is found there generally exists a lateral shift for the outgoing beam through the wedge. We derived a simple expression for calculating this beam shift and interestingly, it provides us a strategy to quantitatively estimate the loss of the wedge material (Im[n]). Additionally, we offered a design of metamaterials, compatible with nano-imprinting-lithography, showing negative refractive index in the visible regime (around yellow-light wavelengths). The multi-layer-system retrieval was utilized to extract the effective refractive index of the metamaterial. It was also intuitively characterized through our wedge setup to demonstrate corresponding phenomena of refraction.

\end{abstract}


\maketitle

Nowadays, metamaterials (MMs) have been well-known in a variety of scientific areas for their versatility in the manipulation of electromagnetic waves, which leads to many intriguing phenomena and unprecedented applications in sub-diffraction imaging, energy resources, ultrafast switches, etc. \cite{Shalaev2007NatPhoton, Soukoulis2007Science, Zheludev2010Science, Boltasseva2011Science, Liu2011CSR, Soukoulis2011NatPhoton, Landy2008PRL, Chen2006Nature}  To flexibly program the designs of MMs for objective applications, the characterization of MMs to obtain effective electromagnetic parameters is very important. Therefore, some related methods have been developed and widely used, such as field average homogenization, \cite{Smith2006JOSAB} S-parameter retrieval, \cite{Smith2002PRB, Smith2005PRE, Koschny2005PRB} and wedge demonstration. \cite{Shelby2001Science, Valentine2008Nature, Chanda2011NatNano} Concerning the demonstration of refraction phenomenon by building a wedge with the designed MM, this provides us with the most intuitive way to judge the property of a negative or positive refractive index. It is even possible to quantitatively determine the effective refractive index of the MM, based on Snell's Law. Since the beginning of the field of MMs, the study of negative index metamaterials (NIMs) has been one of the most important branches in the MMs research because NIMs play a crucial role towards many valuable applications, such as superlensing. Great efforts have been dedicated to realizing low-loss NIMs working in optical even visible regime. \cite{Soukoulis2007Science, Valentine2008Nature, Gacia-Meca2011PRL} Some NIM designs have been proposed at around telecommunication wavelengths. \cite{Valentine2008Nature, Chanda2011NatNano} Quite recently, a NIM in the visible regime was reported. \cite{Gacia-Meca2011PRL} Even though the wedge simulations were provided for some of the optical NIM designs, the reported numerical observations were fairly obscure to distinguish the direction of the refraction. \cite{Valentine2008Nature, Chanda2011NatNano} Here, we show a robust wedge setup, which can unambiguously demonstrate the refraction phenomenon at the interface of MM and ambient material. First, we test our setup with a wedge of a homogeneous NIM and show its nice performance. Then, we apply our wedge settings to several reported optical NIMs and present corresponding demonstrations. Finally, since some of our designed fishnet NIMs working at telecommunication wavelengths (around 1.5 $\mu$m) with nice performance have been successfully produced with the fast and easy nano-imprinting lithography (NIL) technology by our experimental collaborators, \cite{Bergmair2011Nanotech} we made the efforts to push the NIMs to the visible regime and proposed an optimized fishnet design showing negative index around the wavelength of yellow light, which is completely compatible with NIL. With the multi-layer-system retrieval, we extract its effective refractive index. The corresponding refraction phenomena are demonstrated as well through our wedge setup.

Different from the widely adopted wedge setup, in which a wide incident plane wave goes through a small wedge by the MM settled in a relatively large computational region, we consider a fairly large MM wedge under the impinging of a finite beam with analogous distribution of a Gaussian wave (see Fig. 1). With such improvements, the superiority is obvious, i.e., the influence coming from the scattering at the edges of the wedge is minimized and the finite in-coming beam leads to a finite out-going beam through the wedge, which benefits the distinct judge of positive or negative refraction. Our simulations were all performed with well-known commercial software CST Microwave Studio. \cite{CST} We set a finite sized port in front of the MM wedge with a transversal ($x$-direction) electric field and selected the mode showing analogue of gaussian field distribution. The width of the port is about four times the wavelength, which has already efficiently prevented a significant diffraction of itself during the propagation. The MM wedge is settled in an ambient vacuum environment. Our purpose is to demonstrate the refraction phenomenon at the inclined interface between MM and the ambient vacuum. Figures 1(a) and (b), respectively, show the $E_x$ ($x-$component of the electric field) distributions at 400 THz ($\lambda = 750$ nm) through a smooth and step wedge of an ideal homogenous MM with $\epsilon=-1$ and $\mu=-1$. The port size is 3 $\mu$m (4$\lambda$). The size of the wedges in transversal and the longitudinal direction is 9 and 1.5 $\mu$m, respectively, so the angle of the wedge is about 9.5$^\circ$. For the case of step wedge (Fig. 1(b)), the transversal step size is $\lambda/2$. We can see from Fig. 1 very clearly the negative refraction at the inclined interface between MM and the vacuum for both cases. Due to the distinct refraction phenomena with our robust wedge setup, we are also able to get the angle of refraction through the wedge. This allows us to extract the effective refraction index of the MM quantitatively via Snell's Law. 

\begin{figure}[htb]
\centering
\includegraphics[width=7cm]{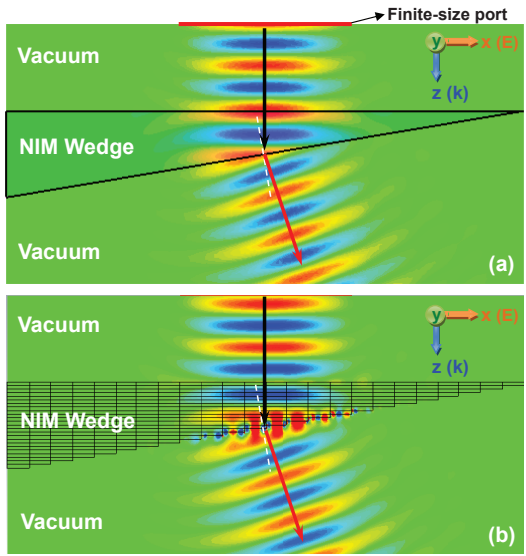}
\caption{(Color online) $E_x$ (x-component of electric field) distribution at $\lambda=750$ nm for (a) smooth wedge and (b) step wedge of homogeneous negative index material ($\epsilon = -1$ and $\mu = -1$), respectively. The port for exciting the incident beam has finite size of 3 $\mu$m wide. The size of the wedge is 9 $\mu$m and 1.5 $\mu$m in the $x-$ and $z-$directions, respectively.}
\end{figure}

Following the simple tests with our wedge setup to the ideal homogenous MM shown above, we definitely would like to see the performance, when realistic MM structures are taken into account. Actually, there have been some reported results of wedge simulations to corresponding optical MMs, for example, Fig. 3(c) in Ref. [14] and Fig. 4(f) in Ref. [15]. However, for the wedge demonstrations shown in Ref. [14-15], it is fairly obscure to distinguish the direction of the refracted beam through the MM wedges, especially in the latter result. Take the structure (fishnet MM) in Ref. [15] as an example. We performed the wedge simulation with our settings at $\lambda=2$ $\mu$m with the result of $E_x$ distribution shown in Fig. 2(a). The geometric parameters of the MM can refer to Ref. [15]. For our MM wedge (see the inset of Fig. 2(a)), in the lateral ($x$) direction, there are 18 unit cells, and a total of 71 layers in longitudinal ($y$) direction at the thicker side. The step size in the $x$-direction is one unit and that in the $y$-direction is $2(t+s)$, where $t$ and $s$ are the thickness of metal and dielectric, respectively. Therefore, the wedge angle is about 10.7$^\circ$. The finite-sized port was applied with a width of 8 unit cells along \textit{x}-direction to generate a finite incident beam. According to Fig. 2(a), negative refraction is clearly demonstrated at the inclined interface between MM and the vacuum, as predicted, based on the retrieval in Ref. [15]. This is an example of NIM working at telecommunication wavelengths. 

We may notice there exists a lateral shift of the refracted beam through the wedge in Fig. 2(a). For this problem, we have completed some theoretical analyses and corresponding simulations. We assume the incident beam has a Gaussian spatial distribution (\textit{E}($x$) $\sim$ exp[$-\frac{x^2}{2\sigma^2}$]) with $\sigma$ denoting the beam width. Under the simplest assumption, by neglecting the dispersion effect at the inclined interface between the wedge and ambient materials, and the multiple reflections in the wedge, we may obtain the first order expression for the lateral shift (along \textit{x-}direction) of the refracted beam (towards the thinner side of the wedge) $d$ $\sim$ $\sigma^2$Im[n]$k_0$tan$\theta$, where Im[n] is the imaginary part of effective refractive index of the wedge material and $\theta$ is the wedge angle. Based on such an expression, it is determined three factors play roles in the occurrence of the beam shift through a wedge: 1) the beam width, 2) the loss of the wedge material, and 3) the wedge angle. We have performed a series of wedge simulations for homogenous NIMs (results not shown here) by changing $\sigma$, Im[n], and $\theta$. The results are consistent with the analytical expression qualitatively. Interestingly, the quantitative characterization of such a lateral beam shift also provides us a strategy to estimate the loss of the wedge material by giving Im[n]. For the case shown in Fig. 2(a), we can get the approximate value of Im[n] as 0.28, which is in good agreement with the retrieval result in Ref. [15].

\begin{figure}[htb]
\centering
\includegraphics[width=7cm]{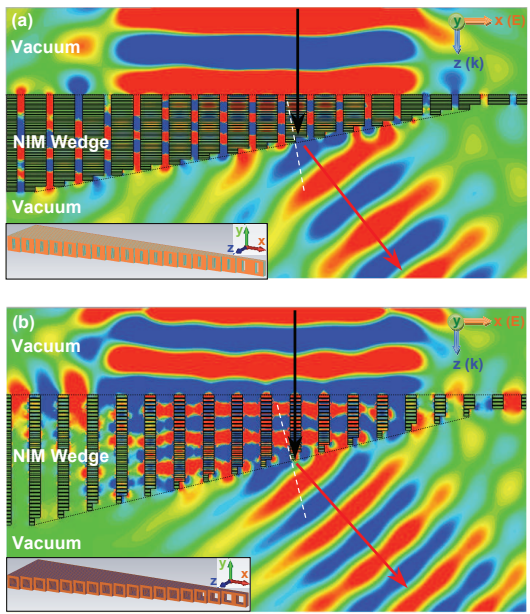}
\caption{(Color online) $E_x$ (x-component of electric field) distributions for wedges of (a) fishnet in Ref. [15] at $\lambda=2$ $\mu$m and (b) Structure 3 in Ref. [16] at $\lambda=750$ nm. Along the lateral ($x$) direction, the sizes of the port, wedge, wedge step are 8-unit, 18-unit, and 1-unit, respectively. Along longitudinal ($z$) direction, the wedge has 71-layer and the step size is $2(t+s)$, where $t$ and $s$ are the thicknesses of metal and dielectric, respectively. The insets schematically show the corresponding MM wedges.}
\end{figure}

\begin{figure}[htb]
\centering
\includegraphics[width=8cm]{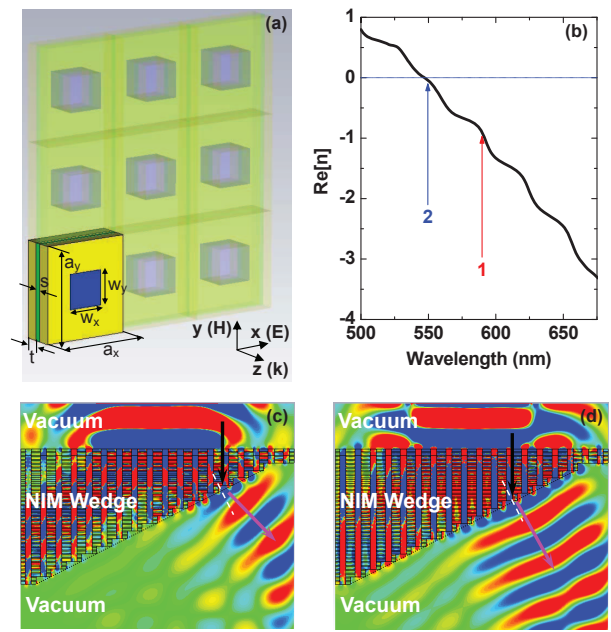}
\caption{(Color online) (a) Schematic of our designed fishnet structure. Materials in yellow, green, and blue are metal (silver), spacer ($\epsilon_d=1.5$) and surrounding material ($\epsilon_s=2.25$), respectively. $a_x=200$ nm, $a_y=220$ nm, $w_x=w_y=80$ nm, $t=35$ nm, and $s=15$ nm. (b) Retrieved real part of refractive index for a 15-layer system of the designed fishnet. (c) and (d) are $E_x$ (x-component of electric field) distributions for the wedge of the fishnet at $\lambda=590$ and 550 nm, respectively. Along lateral ($x$) direction, the sizes of the port, wedge, wedge step are 8-unit, 18-unit, and 1-unit, respectively. Along longitudinal ($z$) direction, the wedge has 71-layer and the step size is $2(t+s)$.}
\end{figure}

Very recently, Garc\'{\i}a-Meca and co-workers designed and fabricated several fishnet MMs with negative refractive index in the visible spectral range. \cite{Gacia-Meca2011PRL} The structures showed a very promising negative index property with good values of figure-of-merit (FOM), which represent low loss NIMs. Even though some retrieval results for the effective electromagnetic parameters of the MMs were shown in Ref. [16], unfortunately they did not perform the wedge simulations to their designs to consolidate the reported negative index property. Here, we take one of these nice fishnet designs (Structure 3 in Ref. [16]) as another example to further examine the performance of our wedge setup. The details of both geometric and materials parameters to such a MM design can be found in Ref. [16]. We applied the same wedge settings as those in Fig. 2(a) with port size (8-unit-cell wide), lateral (\textit{x}) size (18-unit-cell), longitudinal (\textit{z}) size (71-layer) at thicker side, and step size (one unit along $x$ and $2(t+s)$ along $z$). Therefore, the wedge under our study has its angle about 14$^\circ$. The inset of Fig. 2(b) is the sketch of the wedge we built with the MM. Our wedge simulation result ($E_x$ distribution at $\lambda = 750$ nm) is shown in Fig. 2(b) with extremely satisfying performance. The negative refraction is observed distinctly, which confirms the negative index property of the design, as predicted in Ref. [16]. On the other hand, with our derived expression of lateral beam shift through the wedge, the value of Im[n] of the structure can be calculated to be 0.15 approximately.

Based on the tests to homogeneous materials with pre-defined negative index of refraction and realistic optical MMs, our wedge setup has proven to work nicely. It is very robust to demonstrate negative/positive refraction clearly. Hence, it provides a good method to characterize the refractive index of MMs (both Re[n] and Im[n]) qualitatively and even quantitatively.

There has recently been developed a fast and easy NIL-based stacking process towards 3D NIMs on a large area. \cite{Bergmair2011Nanotech}  Since the process is cost efficient and only takes a few seconds, it provides an important strategy for mass production of optical MMs and hence will benefit the realization of various breakthrough applications of MMs as promised. In Ref. [17], our experimental collaborators have successfully produced some of our designed fishnet NIMs at telecommunication wavelengths (around 1.5 $\mu$m) with NIL technique and showed satisfying performance of negative-index property. Afterwards, we have made the efforts to push the NIL compatible NIM designs to the visible regime. We applied the same spacer and surrounding materials, which were adopted in Ref. [17], to our numerical studies, and tried to optimize the geometric parameters of the fishnet structure to achieve good negative-index performance at visible wavelengths. The designed fishnet structure is schematically shown in Fig. 3(a). Silver is taken for the metal layers, due to low loss in the visible regime, \cite{Shen2012PRB, Tassin2012NatPhoton} the spacer between metal layers has $\epsilon_{d}=1.5$, and the surrounding material (hole) is the so-called ormocomp ($\epsilon_{s}$=2.25), which is a good option as adhesion layers to metal. \cite{Bergmair2011Nanotech} Based on our numerical studies, we obtained a design with negative index at wavelengths around 590 nm corresponding to yellow light. The designed fishnet has unit size along $x$ and $y$ directions 200 and 220 nm, respectively, hole size $w_x=w_y=80$ nm, metal layer 35 nm-thick, and spacer layer 15 nm-thick. Via simulations of our multilayered fishnet design with different numbers of layers, we were able to achieve correct retrieval results of effective refractive index from the information of reflection and transmission. \cite{Zhou2009PRB} Figure 3(b) provides the retrieved real part of refractive index for a 15-layer structure. We see negative index response within a quite wide wavelength range (around 600 nm). We also took two sample points at $\lambda=590$ (sample 1) and 550 nm (sample 2), respectively, and applied our wedge setup to the MM to demonstrate the corresponding phenomena of refraction. Figures 3(c) and (d) present the results of wedge simulations at the two sample wavelengths, respectively. At $\lambda=590$ nm, retrieval result predicts Re[n] approximately $-1$. A a clear negative refraction is observed in Fig. 3(c). For $\lambda=550$ nm, the wedge result is also consistent with the retrieval. The beam through the wedge goes almost along the normal of the inclined surface, but with a tiny negative angle of refraction corresponding to refractive index close to 0. Through the estimation of the lateral shifts at two different sampling frequencies, we got the approximate values of Im[n] as 0.52 (at $\lambda = 590$ nm) and 0.30 (at $\lambda = 550$ nm), which are close to the retrieved values of 0.45 and 0.26, respectively.

In conclusion, we provide a robust wedge setup to unambiguously demonstrate refraction phenomenon to characterize the refractive index for metamaterials both qualitatively and quantitatively. The tests for homogeneous materials with pre-defined negative refractive index and realistic metamaterials have proven a superior and satisfying performance of our setup. A simple expression for the lateral shift of outgoing beam through the wedge is derived and provides us a convenient way to estimate the loss of the wedge material. We also offer a nano-imprinting-lithography compatible fishnet design, showing negative refractive index around the wavelengths of yellow light. The design is numerically studied by our wedge setup, which renders consistent results with the multilayer-system retrieval.

Work at Ames Laboratory was supported by the Department of Energy (Basic Energy Sciences, Division of Materials Sciences and Engineering) under contract No. DE-AC02-07CH11358 and by the U.S. Office of Naval Research, Award No. N00014-10-1-0925. This was partially supported by the European Community Project NIM\rule[-2pt]{0.2cm}{0.5pt}NIL (Contract No. 228637) and by ERC Grant No. 320081 (PHOTOMETA).

\bibliographystyle{apsrev}

\newpage

\end{document}